# Towards accurate real-time luminescence thermometry: an automated machine learning approach

Emanuel P. Santos,[a] Roberta S. Pugina,[b] Eloísa G. Hilário,[b] Alyson J. A. Carvalho,[a] Carlos Jacinto,[c] Francisco A. M. G. Rego-Filho,[d] Askery Canabarro,[e] Anderson S. L. Gomes,[a,*] José Maurício A. Caiut,[b] and André L. Moura,[c,†]

[a]Departamento de Física, Universidade Federal de Pernambuco, Recife-Pernambuco, Brazil

[b]Departamento de Química, Faculdade de Filosofia, Ciências e Letras de Ribeirão Preto, Universidade de São Paulo, 14040-901, Ribeirão Preto, São Paulo, Brazil

[c]Instituto de Física, Universidade Federal de Alagoas, Maceió-AL, Brazil

[d]Campus Coruripe, Instituto Federal de Alagoas, Coruripe-Alagoas, Brazil

[e]Grupo de Física da Matéria Condensada, Núcleo de Ciências Exatas – NCEx, Campus Arapiraca, Universidade Federal de Alagoas, Arapiraca-AL, Brazil

*Corresponding author. E-mail: andersonslgomes@gmail.com

†Corresponding author. E-mail: andre.moura@fis.ufal.br

**Abstract**

Luminescence thermometry has been extensively exploited in the last decades both from the fundamental and applied point of views. The application of photoluminescent nanoparticles on the microscopic level based on rare-earth doped (RED) nanostructures is yet a challenge. Distinct underlying physical mechanisms in the RED nanomaterials have been exploited, such as intensity ratio between radiative transitions associated with thermally coupled energy levels, energy peak and lifetime of an excited state variations with the temperature. The drawbacks of such systems are the relatively low thermal sensitivity ($S_r$), and the large temperature uncertainty. To overcome that, several research groups have been seeking new functionalized materials. The majority of the efforts have been directed towards increasing $S_r$ with record around 10 %°C$^{-1}$, which is, however, considered unsatisfactory. We propose the use of an automated machine learning tool to retrieve an ideal pipeline improving the response of photoluminescence thermometers. As a proof-of-concept, we used Nd$^{3+}$-doped YAG nanoparticles, excited at 760 nm, and the photoluminescence spectra in the range from 860 nm to 960 nm as input parameters. In addition to the improvement in the accuracy (> 5.5× over traditional methods), the implementation is very simple, without the requirement of any deconvolution procedure or knowledge of any underlying physical mechanism. Our findings demonstrate that this approach is resilient to natural variances across various spectral acquisitions, which may otherwise lead to an inaccurate estimation of temperature, opening the door for real-time applications. Our open-source code is designed to be accessible to individuals without programming experience.

**INTRODUCTION**

Accurately determining temperature is of paramount importance, as it is a fundamental parameter in both scientific and technological research, besides practical applications. Various techniques can be utilized to accomplish this, such as liquid thermometers, thermocouples, and blackbody radiation analyses. Each method works satisfactorily on specific application. Among such methods, the luminescence thermometry, based on trivalent rare-earth ($RE^{3+}$) ions, has been extensively investigated and can be utilized for remote sensing in both macroscopic and microscopic scales.[1,2] For macroscopic applications, there are commercial options available. However, to employ this technique on a microscopic scale, such as in micro-electronics, catalysis, microfluidics, and biologic systems, the development of photoluminescent nanomaterials is necessary. Many previous investigations were focused on the photoluminescence of thermally-coupled energy levels of $RE^{3+}$ ions, which were conducted in both bulk[3] and nanoscale[4] levels. The underlying physical mechanism is the relative population of energy levels, which is governed by the Boltzmann distribution. In fact, alterations in the photoluminescence spectra resulting from temperature changes can be utilized to calibrate and operate an optical thermometer. Remarkably, the intensity ratio between two emission transitions (named as Luminescence Intensity Ratio, LIR) has been pointed out as the most promising method because it can, in principle, deal with practical drawbacks, such as intrinsic random fluctuation of the excitation beam intensity and other downsides.[5] The effectiveness of these thermometers is evaluated by the relative thermal sensitivity, which is defined as $S_r = 100\% \times R^{-1}dR/dT$, where R is the intensity ratio between two emission transitions, and T denotes the temperature. Therefore, several studies focus on enhancing $S_r$ in order to make the thermometers more widely applicable. Typical values of $S_r$ for $RE^{3+}$ are about 1 %$C^{-1}$, but some unprecedented approaches demonstrated values around 10 %$C^{-1}$ by exploiting different physical mechanisms.[4,6–8] For example, da Silva et al.[6] considered the emission of $RE^{3+}$ (thulium) and the second-harmonic generation in the interaction of the laser beam with the host medium ($NaNbO_3$), demonstrating values as large as 12 %$C^{-1}$; Ximendes et al., using core-shell engineering to improve $S_r$ ($\approx$ 5 %°$C^{-1}$) obtained *in vivo* 2D subcutaneous thermal video;[8] and Xu et. al., used a hybrid system based on triplet sensitized upconversion nanoparticles to develop a nanosensor with a high thermal sensitivity of $\approx$ 7.1% °$C^{-1}$.[4]

In theory, increasing the value of $S_r$ could address several of the issues and drawbacks associated with luminescent nanothermometers. This is because a high $S_r$ value would result in significant changes in R with respect to T, thereby enabling more precise temperature measurements. In this context, numerous studies have been performed, but it appears that exceeding 10 %C$^{-1}$ remains a significant challenge. Furthermore, as evidenced by several sets of published experimental data, significant fluctuations in R with respect to T can prevent a unique correspondence between these quantities. The accuracy of the nanothermometer is usually measured by the temperature uncertainty δT = $S_r^{-1}$δR/R, where δR/R is the relative uncertainty in measuring R, which is limited by the acquisition system. Notice that δT depends inversely on $S_r$. The best reported values of δT are between 0.1 and 0.3 K.[9,10] Another way to determine the accuracy of an optical thermometer is by computing $σ_T$, which is the standard deviation of the difference between the calculated temperatures and the reference ones.[11] Besides large values of $S_r$, neither δT or $σ_T$ are computed in several reported works, but one can see large deviations of the calculated temperature with respect to the target one. Thus, large temperature uncertainties make unfeasible the operation of the nanothermometer, specially for biological applications that require δT ≈ 0.1 °C.[12] As demonstrated by van Swieten et al.,[13] intrinsic characteristics of the detection system (noise and background) and/or experimental conditions can impact the precise operation of luminescent thermometers, specially $σ_T$. In general, the spectra acquisition is performed with spectrometers equipped with charge-coupled device (CCD), and the signal-to-noise ratio can be maximized by increasing the integration time. This process is equivalent to sum (or average) a series of spectra, and the noise, as a sum of random variables, is minimized unveiling clear spectra. However, in dynamic processes, in which the temperature presents fast variations with time, working with large integration times is unfeasible.

CCD-based spectrometers can acquire data in wide spectral ranges, for example, from 330 nm to 1180 nm. Then, using only the peak intensity of two given transitions to operate ratiometric thermometers discards a lot of information available in the spectra, such as peak positions and bandwidths. In a remarkable approach, Maturi et al.[14] applied the mathematical method of Multiple Linear Regression (MLR) considering as temperature-dependent parameters, those ones which changed linearly with T. As a consequence, a record $S_r$ of 50 %°C$^{-1}$ was obtained. Although the authors considered only parameters that varied linearly with T, nonlinear dependences could also be taken into account with a generalized method for multiple nonlinear regression, whose computational implementation with built-in functions could be performed with operations already present in several programming platforms (e.g. Python and Matlab).

Machine Learning (ML) is a cutting-edge approach utilized in several scientific data-driven fields. Essentially, computational models have the capability to learn and enhance their performance as they process more data.[15] ML encompasses several techniques such as supervised learning (e.g., regression, classification), unsupervised learning (e.g., clustering, dimensionality reduction), semi-supervised learning, reinforcement learning, and deep learning (a subset of neural networks). Supervised learning involves training models using labeled data to perform specific tasks such as classification, where labels are assigned to new data points, and regression, where continuous values are predicted. The current paper focused on the task of regression, which aimed to predict temperature values (the target) based on spectral data as input features.

There are various supervised learning algorithms that can be used for regression tasks, such as Linear Regression, Decision Trees, Support Vector Machines, Gradient Boosting, and Neural Networks, to cite a few. However, selecting the best algorithm and optimizing its hyperparameters can be a challenging, time-consuming and technical process. Automated ML (AutoML) tools, such as TPOT (Tree-based Pipeline Optimization Tool),[16] which is an open-source Python library, can help overcome these challenges by automatically exploring a wide range of models and hyperparameters to identify the optimal solution for a given problem via genetic programming. In this study, TPOT was utilized to simplify the process of selecting the most effective regression model for predicting temperature values from spectral data, retrieving an optimized pipeline. In essence, AutoML also helps democratize ML among non-technical researchers by simplifying the process of building and selecting the best ML pipelines, and have already been successfully applied within scientific domains.[17,18]

One can wonder about the differences between a ML approach and a canonical algorithm. The main difference is that in the first we insert inputs and answers to retrieve rules, and in the second inputs and rules are given in order to get answers. Interestingly, ML has been used successfully in physics, even in situations where there is no clear underlying physical mechanism. This is because ML can promptly identify patterns and correlations in large datasets. While traditional physics approaches rely on well-defined equations and principles, ML provides a more data-driven approach that can be useful in identifying new phenomena and making predictions in complex systems. Heuristically, using general-purpose learning procedures, ML can identify phase transitions,[19] verify correlations,[20,21] help in designing photonic structures,[22] and devices,[23] thus accelerating technology developments.[24] There are reports that used some ML approaches to thermometry, such as Dimensionality Reduction,[25] Single Value Decomposition,[26] and Neural Networks to improve the response of luminescent thermometers.[27–35] But, the Neural Networks were first exploited for luminescence thermometry (rhodamine B) in 2014.[27] As pointed out by the authors, there are several possibilities to consider as input parameters: the bare fluorescence data or combinations of spectral features, e.g., peak/integrated intensity, peak wavelength, and a selection of values of the normalized spectrum. Since then, several reports appeared in the literature,[27–35] exploiting, for example, the photoluminescence properties of rhodamine B,[28] quantum dots,[29–31] trivalent europium ions,[32], and $Y_3Al_5O_{12}:Cr^{3+}$ phosphor.[35] In general, those works report enhancements of the luminescent thermometer accuracy when compared to the traditional luminescence intensity ration (LIR) and even the multiple linear regression (MLR).[35] The calibration procedure involves getting a series of spectra under fixed temperatures. Part of the acquired data can be randomly sorted to test the validity of the procedure or it can be performed by using an independent set of measurements indicating the robustness of the methodology.[35]

Here, we investigate intrinsic fluctuations in CCD-based spectrometers from one spectral acquisition to the other, showing that they can lead to large uncertainty in the temperature determination for the traditional methods of LIR and MLR. Afterwards, we investigate ML approaches to demonstrate the minimization of temperature uncertainty even for "bad" or limited data sets. This work paves the way for real-time optical thermometry, since the methodology can be applied directly to several systems, i.e., it suffices to train the machine by inputting a given number of spectra and the corresponding temperature. Given the simplicity of the method, we make available an open-source code, which can be used for several photoluminescent materials and different spectral ranges.

**METHODS**

A. NANOPARTICLES PREPARATION AND CHARACTERIZATION

The $Nd^{3+}$:YAG (3.5%) particles were prepared by spray pyrolysis using as a precursor 50 mL of boehmite suspension (0.2 M) doped with 60% of $Y^{3+}$ ion (mol/mol in relation to $Al^{3+}$ from boehmite composition) and co-doped with 3.5% $Nd^{3+}$ (mol/mol in relation to $Y^{3+}$), as previously described.[44] For this, 11.3 ml of $Y(NO_3)_3$ (0.5 M) and 3.5 ml of $Nd(NO_3)_3$ solutions (0.1 M) were added to 25 ml of ultrapure water (T= 83 °C). In the sequence, 2.44 g (0.01 mol) of the aluminum tri-sec-butoxide was added and hydrolyzed for 1 h under stirring. Finally, 1.0 ml of nitric acid as a peptizing agent was added. After cooling the suspension, the volume was adjusted to 50 mL with ultrapure water. The suspension was spray pyrolyzed into the spray pyrolysis system. The collected powder was heated at 1100 °C for 12 h.

The particles were structurally and morphologically characterized by X-ray diffractogram and scanning electronic microscopy. The luminescent properties were studied by photoluminescence spectroscopy.

B. OPTICAL EXPERIMENTS

In the experimental procedure, starting at room temperature (≈ 30 °C), we turned on/off the heating source rising and lowering the particles temperature, which was measured with the thermal camera for calibration purposes (FLIR E40, temperature detection range from -20 to 650 °C with accuracy of ± 2 °C). During this procedure, the particles (pressed gently in a metallic sample holder) were excited at 760 nm with a tunable (720 nm – 850 nm) Ti:Sapphire (Ti:Al$_2$O$_3$) laser at a constant power (70 mW), and photoluminescence spectra were recorded, which allowed the association between each spectra and the particles temperature. The procedure of rising/lowering the particles temperature was repeated, ensuring reproducibility of the experimental results. The spectrometer consists of a diffraction grating and a CCD. In the interest wavelength range (860 – 960 nm) there are 100 pixels that, combined with the optical path length and input diffraction grating, provided a spectral resolution of 2.4 nm.

C. AutoML

We can then create an instance of TPOT and fit it to our data. TPOT will then automatically search through a range of machine learning models and pipelines to find the best combination for our problem. TPOT uses genetic algorithm and some of the main parameters are:

    1. generations: the quantity of generations that TPOT should run for. Each generation involves evolving new pipelines and evaluating their fitness on the dataset.

    2. population size: the number of pipelines to keep in each generation. Larger populations may lead to better performance, but will also increase the time and computational resources required.

    3. cv: the cross-validation strategy to use when evaluating pipelines.

    4. scoring: the metric to use when evaluating pipelines.

The larger the population and generation sizes, the longer TPOT takes to complete. After running TPOT, we can use and save the best pipeline to make predictions on new data. We can also export the pipeline to a Python script or pickle file for later use. Please, check our repository for further technical details.[43]

D. DATA PRE-PROCESSING

For the Machine Learning approach, no data pre-processing was performed in the sense that we did not realize any deconvolution of curve fitting. However, given the large spectral range acquired by the spectrometer, we restrict the spectra to the wavelength range of interest (860 nm – 960 nm).

*Santos et al. Towards accurate real-time luminescence thermometry: an automated machine learning approach*

**RESULTS AND DISCUSSION**

The YAG phase of the crystalline particles (see methods section for preparation details) was confirmed by X-ray diffractogram (Fig. 1a) with the observation of peaks associated to the structural planes of YAG. The photoluminescence spectrum (Fig. 1b) exhibits well-resolved and narrow bands characteristic of the $Nd^{3+}$ *f-f* transitions in the YAG crystalline structure. Spherical particles were observed by scanning electron microscopy (inset of Fig. 1b).

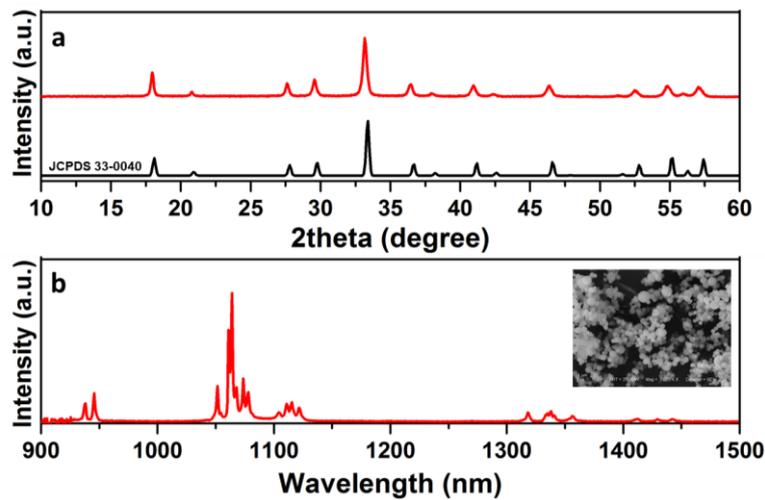

**Fig. 1. a** X-ray diffraction of the $Nd^{3+}$:YAG particles (red line) compared to the pattern for the YAG phase (black line). **b** Emission spectrum of the $Nd^{3+}$:YAG particles and (insert) its spherical morphology obtained by scanning electron microscopy.

The resonant ground-state to excited state transition of $Nd^{3+}$ at 760 nm ($^4I_{9/2} \rightarrow \{^4F_{7/2}, ^2S_{3/2}\}$) provides emission in a broad spectral range by down-conversion processes, as represented in the $Nd^{3+}$ partial energy level diagram in Fig. 2. The $Nd^{3+}$ at the $\{^4F_{7/2}, ^4S_{3/2}\}$ states relax to the lower-lying levels $\{^4F_{5/2}, ^2H_{9/2}\}$ and subsequently to the metastable $^4F_{3/2}$ state with phonon emissions. From the $^4F_{3/2}$ level, the $Nd^{3+}$ ions can relax nonradiatively to the lower-lying level $^4I_{15/2}$ with phonon emissions, or radiatively to the $^4I_{15/2}$ (1750 nm), $^4I_{13/2}$ (1320 nm), $^4I_{11/2}$ (1064 nm), and $^4I_{9/2}$ (880 nm) states. Also, there are the possibilities of energy transfer among $Nd^{3+}$-$Nd^{3+}$ pairs producing upconversion emission [$^4F_{3/2}, ^4F_{3/2} \rightarrow \{^4G_{7/2}, ^2G_{3/2}\}, ^4I_{13/2}$] or phonon emissions [$^4F_{3/2}, ^4F_{3/2} \rightarrow ^4I_{15/2}, ^4I_{15/2} \rightarrow ^4I_{13/2}, ^4I_{13/2} \rightarrow ^4I_{11/2}, ^4I_{11/2} \rightarrow ^4I_{9/2}, ^4I_{9/2}$]. However, these processes are not relevant here. Rising the particles temperature affect the nonradiative relaxation rates as well as promotes population redistribution among thermally-coupled energy levels of the $Nd^{3+}$ due to the increase in the phonon occupation number. We will focus on the $Nd^{3+}$ radiative transition $^4F_{3/2} \rightarrow ^4I_{9/2}$ with emission at around 880 nm. The number of Stark levels in each manifold is 2 and 5 for the $^4F_{3/2}$, and $^4I_{9/2}$ states, respectively. Accordingly, several peaks are observed for the transitions at around 880 nm (Figs. 2 and 3a), even with the low spectral resolution (2.4 nm) of the acquisition system. One can also see emission peaks at around 940 nm, which are associated to $\{^4F_{5/2}, {}^2H_{9/2}\} \rightarrow {}^4I_{11/2}$ transitions. Then, rising the temperature can decrease the emission intensities due to nonradiative relaxations from the $^4F_{3/2}$ and $\{^4F_{5/2}, {}^2H_{9/2}\}$ states, promoting population redistribution among the Stark levels of each manifold affecting the peak wavelengths, bandwidths, and relative intensities, whose spectral changes are not so clear in Fig. 3a due to the low spectral resolution. Another pathway to decrease the emissions whose electronic transitions originate from the $^4F_{3/2}$ state is by ladder-thermal excitation to upperlying level $\{^4F_{5/2}, {}^2H_{9/2}\} \rightarrow \{^4F_{7/2}, {}^4S_{3/2}\} \rightarrow {}^4F_{9/2}$, which is relevant only at high temperatures.[36–38]

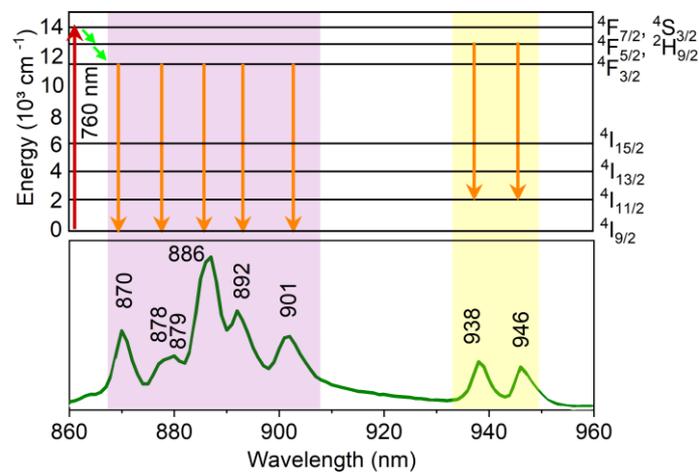

**Fig. 2.** $Nd^{3+}$ partial energy level diagram illustrating the excitation (760 nm) and relaxation pathways. The bottom part of the figure shows the corresponding $Nd^{3+}$:YAG emission spectrum.

Figure 3 shows the experimental temperature measurements carried out with the $Nd^{3+}$ nanoparticles. Given the relative changes among the eight intensity peaks in Fig. 3a (marked by red dots), it is reasonable to calibrate a thermometer based on the intensity ratio of two peaks. There are 28 possibilities for that (8C2), and we tested all of them. At first sight, the signal-to-noise ratio in the spectra in Fig. 3a looks good because no noise traces are seen. Despite of this, the relationship between the intensities at two given wavelengths presents large fluctuations due to the intrinsic noise in the detection system from one acquired spectrum to the other (Fig. 3b for the intensities at 946.0 nm and 879.0 nm – the solid curve is a linear fitting only to guide the eyes). The correlation of those quantities was evaluated by the Pearson coefficient (a measure of linear correlation between two quantities), which depends on the pair of wavelengths chosen. For the particular case in Fig. 3b, the Pearson coefficient was 0.98 indicating a good correlation among the quantities, but one can see large dispersion around the linear fitting that is due to the noise of the detection system from one spectrum to the other.

The temporal evolution of the intensity in each peak (879.0 nm and 946.0 nm) are represented in Figs. 3c and 3d, respectively. Together with the noise, characterized by random fluctuations from the intensity in one spectrum relative to the other, there are slow variations and abrupt changes. The noise is associated to the detection system (spectrophotometer), while the last two to the excitation laser. The variations due to the excitation laser are, in principle, eliminated by making the intensity ratio at the two given wavelengths (ratiometric thermometers). In fact, both problems (slow and abrupt variations) were eliminated, and the intensity ratio presents only the noise of the detection system (Fig. 3e). By associating each intensity ratio (R) to the corresponding particles temperature (Fig. S1 in the Supporting Information, SI), we calibrate the optical thermometer by fitting a linear function to the data (R = 0.94 + 0.0013×T). Even knowing that the Boltzmann distribution is given by $\exp[-\Delta E/(k_B T)]$, the dependence may seem linear (pseudo-linear behavior) because the exploited range of temperature is relatively low.

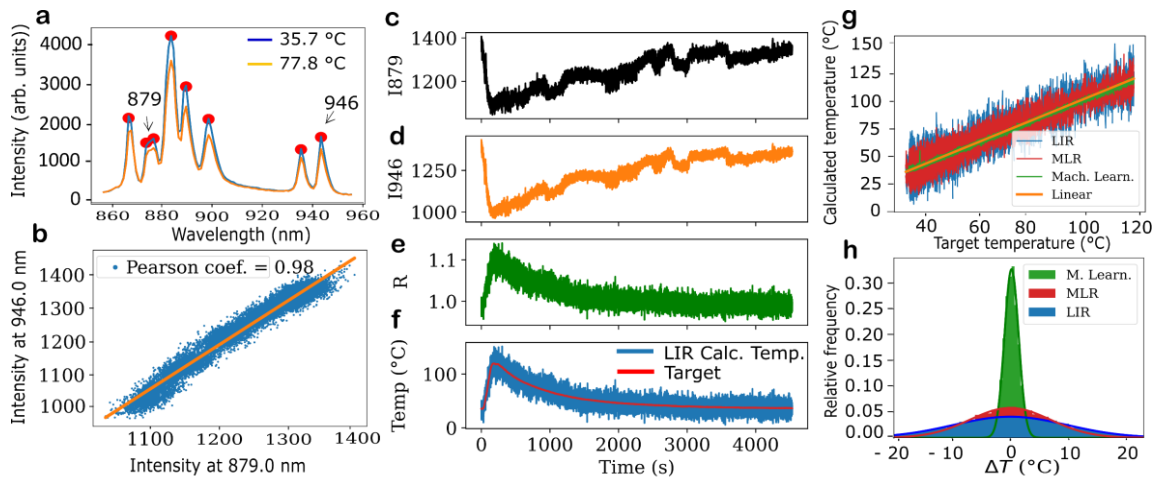

**Fig. 3. a** $Nd^{3+}$ photoluminescence spectra due to the $^4F_{3/2} \rightarrow {}^4I_{9/2}$, and $\{^4F_{5/2}, {}^2H_{9/2}\} \rightarrow {}^4I_{9/2}$ electronic transition under the indicated temperatures. The red dots are the peak wavelengths from which the corresponding intensities were used to calibrate ratiometric thermometers. **b** The relationship between two given wavelengths, which is not well-defined due to intrinsic fluctuations of the acquisition system from one acquired spectrum to the other. The solid line is a linear fit to the data. **c** and **d** Intensity at the two indicated wavelengths as a function of time in the process of increasing and decreasing the particles temperature. **e** Intensity ratio (R) for the emissions at 879.0 nm and 946.0 nm. Not shown, but the representation of R × the target temperature (measured with the thermal camera) allowed the calibration of a ratiometric thermometer. **f** Calculated temperature from R and target temperature. **g** Calculated temperature from different methods [Luminescence Intensity Ratio (LIR), Multiple Linear Regression (MLR), and Machine Learning (ML)] versus the target temperature. The solid line is the expected outcome (Target temperature × Target temperature). **h** Temperature uncertainty when using each method in g, measured by the probability distribution of the calculated minus the target temperature. The solid lines represent Gaussian fittings to the data.

We checked the optical thermometer validity by representing the calculated temperature, and the measured by the thermal camera (target temperature, solid line in Fig. 3f) as a function of time in the heating and cooling process. At first glance, the results may sound

*Santos et al. Towards accurate real-time luminescence thermometry: an automated machine learning approach*

satisfactorily, since the behaviors of heating/cooling were recovered. However, the high fluctuations of the calculated temperature (Fig. 3f) indicate the low precision of the optical thermometer, i.e., large uncertainty are present in the measurements. In order to evaluate the uncertainty, the calculated temperature ($C_T$) as a function of the target temperature ($T_T$, measured with the thermal camera) is represented in Fig. 3g. We considered $\sigma_T = [\sum(C_T-T_T)^2/N]^{1/2}$ as a figure of merit, where N is the number of spectra over which the sum extends. Also, we represented a histogram of $C_T - T_T$ in Fig. 3h. The probability distribution could be fitted by a normalized Gaussian function whose standard deviation coincides with $\sigma_T$. For the LIR data in Fig. 3h, an unsatisfactory $\sigma_T$ = 9.8 °C was obtained. At this point, one should notice that, in several published data, the performance of optical thermometers is evaluated by the relative thermal sensitivity ($S_r$) as well as the thermal resolution ($\delta T$), which were defined above. From LIR data (Fig. 3g), $S_r$ = 0.13 %°C$^{-1}$ (maximum) and $\delta T$ = 9.7 °C (minimum) at 30 °C. The $S_r$ is compatible with previous works on Nd$^{3+}$ based thermometers.[39–41] However, the thermometer underestimates the temperature (large $\delta T$), which makes unfeasible the operation of the optical thermometer for several applications.

     Given the 28 possibilities of choosing two given peaks in the Nd$^{3+}$ photoluminescence spectrum to calibrate the ratiometric thermometer (Fig. 3a), one can wonder of using the multiple linear regression.[14] First, we considered all 28 possibilities but, as described in the SI (Fig. S2), some of them presented enormous fluctuations worsening the outcome, i.e., $\sigma_T$ raised from 6.9 °C to 22.8 °C (Fig. S3 in the SI). Then, we established a threshold for combination of pairs whose dependence of the Calculated Temperature × Target Temperature gave a Pearson coefficient ≥ 0.85. The corresponding 6 pairs were the intensities at 877.0 nm and 908.0 nm, 879.0 nm and 946.0 nm, 878.0 nm and 901.0 nm, 877.0 nm and 946.0 nm, 938.0 nm and 901.0 nm, and 938.0 nm and 946.0 nm, respectively. The calibrations of the optical thermometers were performed by plotting the target temperature as a function of the intensity ratio $R_i$ (i = 1, 2, …, 6). The linear fitting of the data leads to linear equations $T = \beta_{0,i} + \beta_i R_i$. The maximum $S_r$ of each thermometer was 0.11, 0.13, 0.12, 0.13, 0.11, and 0.13 %°C$^{-1}$, respectively. To perform the multiple linear regression, the temperature was written as a function of the intensity ratios $T(R_1, R_2, …, R_6) = \beta_0 + \beta_1 R_1 + \beta_2 R_2 + … + \beta_5 R_5$, with $\beta_0 = \sum \beta_{0,i}$. $S_r$ can be written as $[\sum(\beta_i R_i)^{-2}]^{1/2}$,[14] which in the present case becomes 0.30 %°C$^{-1}$, i.e., 2.3-fold larger than the obtained by single fit of intensity ratio. Mathematically, one has an improvement of $S_r$, and the noise is lowered considerably thanks to the sum of noise (random variables) in the multiple dependences (Fig. 3g). Accordingly, there is an improvement in $\sigma_T$ that decreased from 9.8 °C (Figs. 3g and h) to 6.9 °C, and $\delta T = [S_r^{-1}(\sum(\delta R_i/R_i)^2)]^{-1}$ [14] decreased from 9.7 °C to 8.7 °C, but it is still unsatisfactory. It is worth mentioning that in Ref. [14] the authors exploited the photophysical properties of

green fluorescent protein (Protein Data Bank ID 2y0g), which, after spectral deconvolution of two peaks, unveiled the impact of temperature on the intensity ratio of integrated areas, energy peak positions, and full width at half maximum of the two peaks. Considering the multiple linear regression, $S_r$ = 3.0 %°C$^{-1}$ was determined, which is tenfold larger than the highest value obtained in single parametric sensing (0.33 %°C$^{-1}$). Also, the authors investigated inorganic $AgS_2$ nanoparticles, which is known to present large photothermal changes.[42] They exploited the peak wavelength and the intensity ratio at two wavelengths as thermo-responsive parameters. As a consequence, giant values of $S_r$ (50 – 23 %K$^{-1}$) and $\delta T$ (0.025 – 0.33 K) were obtained for temperatures varying from ≈ 295 to ≈ 344 K. Notice that the realization of spectral deconvolution is not practical to be implemented, and for the case of Nd:YAG particles would require the use of high resolution spectrometer in order to resolve the overlapped emission bands.

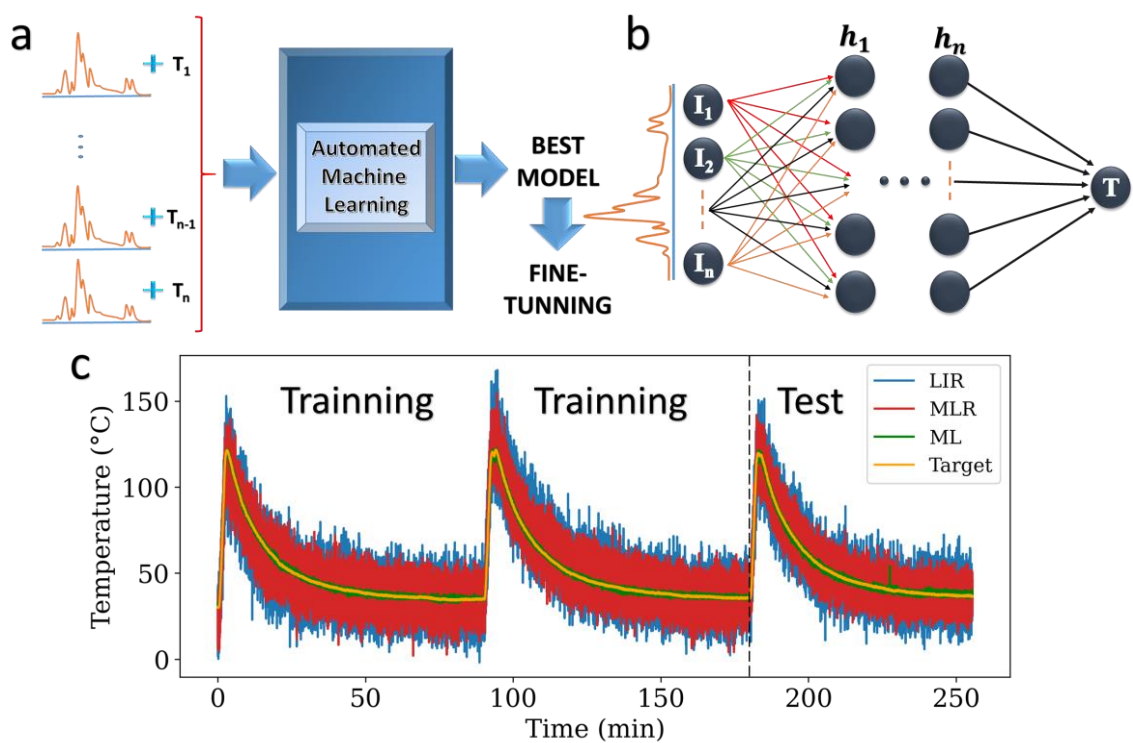

**Fig. 4.** Deep Learning approach to optical thermometry. **a** Representation of the process: the input spectra is discretized in bins of 1 nm. Each bin works as an input neuron and the temperature as the output neuron. Between the input and output, one can use several Machine Learning tools. **b** In the Neural Network, which is a Deep Learning approach, a multilayer perceptron is created by including hidden layers between the input and output. **c** Considering three cycles of rise and lower the particles temperature, the first two were used as training, while the third was used as testing.

In the next step, we investigate the use of ML framework to improve the precision of the luminescent thermometer. We fed TPOT, the AutoML tool, with the spectral data and their corresponding temperature to automatically search through different ML models and pipelines to find the best combination for our problem (Fig. 4a). After identifying the best pipeline proposed by TPOT, we may further fine-tune it to improve its performance on our specific dataset. We provide a repository with Jupyter notebooks and Python scripts that can easily be used for the reproduction of both the AutoML and ML results in Ref. [43]. The methods section provides further technical details on the implementation of TPOT.

According to our TPOT results, the MLP (Multilayer Perceptron) algorithm was identified as the best approach for our task. MLP is a type of neural network, which is a ML model inspired by the structure and function of the human brain. In a neural network, one creates layers of interconnected nodes (called neurons) that process information. The first layer is called input layer, which receives the data that we want to analyze, the spectral data in our case (Fig. 4b). The final layer is the output layer, which produces the predictions of the temperature. In between the input and output layers, we can include one or more hidden layers that help the network learn complex relationships in the data. Each neuron in a neural network has a set of weights that it uses to combine the inputs from the previous layer. Then activation function is applied to the weighted sum of inputs to produce its output. The activation function is typically a non-linear function that helps the network learn complex relationships between the inputs and outputs. It is a feed-forward process because of the one-way flow from the input to the output layer.

Remarkably, while a simple linear fit contains only two parameters to be determined with data, the MLP requires several parameters (depending on the size and number of hidden layers). Thanks to that, the procedure can capture several photothermal changes from the input spectra without any deconvolution process, i.e., it suffices to label each spectrum to the corresponding particles temperature (this labeling process characterizes the learning process as supervised). The cost is that, as many hidden layers are created, many spectra and the associated temperature are required in order to properly train the machine, i.e., determine the coefficients. During the fine-tuning stage of our model, we experimented with different configurations of hidden layers. Our findings showed that using three hidden layers, each with 50 neurons, resulted in the best performance compared to using three layers with 100 neurons each, as suggested by TPOT. We also tested other layer configurations, ranging from one to three layers, with 50, 100, 200, 500, or 1000 neurons each. A table (S1) comparing various metrics related to these ML experiments is provided in the Supporting Information. It is worth emphasizing the supervised ML creates universal rule that maps inputs to outputs. For luminescent thermometry, the algorithm should try to minimize the squared error between the calculated (prediction) and the reference (target) temperatures iteratively from one training spectrum to the other in order to map input spectra to a continuous output set of temperatures.

Instead of applying any deconvolution to the $Nd^{3+}$ photoluminescence spectra or choosing some pair of wavelengths to make a ratiometric thermometer, we used the $Nd^{3+}$ entire spectra (860 – 960 nm) as input parameters, which contain all information about photothermal changes limited by the low spectral resolution of our data. In analogy with a neural network, each of the 100 pixels from the CCD coupled to the spectrometer (measures the intensity in a wavelength range of 1 nm) is an input neuron (Fig. 4b). The objective is to map each spectrum to the particle's temperature, i.e., starting from the input neurons (intensity at each CCD pixel) calculate the corresponding particles temperature (unique output neuron). As a common practice in ML framework, we divided the experimental data in two groups: training and testing. The Machine is supposed to learn with the training data, while its validity is checked with the test ones.

We considered three cycles of rising/lowering the particles temperature (Fig. 4c). The first two runs (rising and lowering the temperature) were used to training, while the last cycle was used for testing. In this way, the robustness of the method can be appreciated since the test is over different data from those used in the calibration (training). As part of a cross-validation demonstration, we also use the training set to predict the temperature. This allows us to evaluate the model's performance on both seen (training set) and unseen (test set) data points, which is crucial for assessing the generalization power of the ML model. The predictions made with ML are much better than the conventional methods described above (LIR and MLR) (Fig. 4c). In order to make a quantitative comparison, we represented the probability distribution of $C_T - T_T$ in Fig. 3h (the third cycle in Fig. 4c corresponds to the experimental data in Fig. 3). The narrower distribution for ML signalizes the superior performance over the traditional LIR method and even the MLR. From the Gaussian fittings of the data in Fig. 3h, $\sigma_T$ lowered from 9.8 °C (LIR) and 6.9 °C (MLR) to 1.2 °C using ML. Then, an expressive improvement (> 5.5×) in $\sigma_T$ was obtained, and can be a new paradigm in luminescence thermometry, given the simplicity of the method, i.e., does not require any curve fitting or deconvolution process in the calibration procedure. Furthermore, it is noteworthy that no data pre-processing was performed to cope with the intrinsic fluctuations of the excitation laser intensity, i.e., the ML method deals with that naturally. Notice that, $\sigma_T$ can be further reduced by using higher resolution spectrometers (sub-nanometers) from which more photothermal spectral changes can be unveiled. Additionally, there is the uncertainty of the thermal camera that was used as reference. Although it has a nominal resolution of ±2 °C, we observed fluctuation on the order of 1 °C when measuring the particles temperature under the same experimental conditions (Fig. S4 in the SI). This large fluctuation of the reference temperature inhibits a unique correspondence between spectrum and particles temperature making the ML outcome less precise. In spite of that, the ML method presents a temperature uncertainty very close to the resolution of the calibration system. It is worth mentioning that we are using $\sigma_T$ to compare the ML with the LIR and MLR methods because it is not possible to define $S_r$ neither $\delta T$ for ML.

**CONCLUSION**

We investigated a pragmatic approach to luminescence thermometry by exploiting ML methods using, as a proof-of-concept, the photoluminescence spectra from trivalent neodymium ions ($Nd^{3+}$ in YAG submicrometric particles) in the range 860 nm – 960 nm, which corresponds to the $Nd^{3+}$ electronic transitions $^4F_{3/2} \rightarrow {}^4I_{9/2}$ and $\{^4F_{5/2}, {}^2H_{9/2}\} \rightarrow {}^4I_{9/2}$. The advantages of using ML over traditional methods, such as Luminescence Intensity Ratio and even the recently approach of Multiple Linear Regression,[14] is that the ML requires only labeled spectra as input parameters, i.e., in the supervised learning, the training process requires the association of spectra and the corresponding particles temperature without any deconvolution procedure. Within the ML toolbox, we tested over 37 algorithms and the Neural Network provided the best results. The temperature uncertainty (1.2 °C) using ML was very close do the resolution of the thermal camera (2.0 °C), taken as the target ones. We foresee that the ML based luminescent thermometer precision can be enhanced by using high resolution spectrometers able to unveil all the photothermal spectral characteristics providing accurate thermal readouts as well as by using more precise thermometers in the calibration procedure. Other luminescent systems should also be investigated in order to find the best luminescent thermometer. We demonstrated the robustness of the method by testing over a different data set used for training. The method is also able to cope with intrinsic intensity variations of the excitation laser. The code is available online and can be used for people without previous knowledge in programming. For real-life applications, the ML algorithm can be integrated into the data acquisition system in order to measure temperature in real time. Currently, the richest platform for ML is PYTHON programming language, and there are several development environments, such as SPYDER, which can integrate data acquisition and fast post-processing. Then, the integration of trained ML and data acquisition can result in real-time optical thermometry with high accuracy.

**SUPPORTING INFORMATION**

Luminescence intensity ratio for the 28 possibilities in Fig. 3. Temperature measurement using the multiple linear regression method considering the 28 possibilities for the intensity ratio, and stablishing a threshold with Pearson coefficient. Fine-tuning of the best machine learning architecture. Uncertainty in the temperature measurements with the thermal camera.

**ACKNOWLEDGEMENTS**

We acknowledge financial support from the Brazilian Agencies: Fundação de Amparo à Pesquisa do Estado de Alagoas (FAPEAL), Coordenação de Aperfeiçoamento de Pessoal de Nivel Superior (CAPES) – Finance Code 001, Fundação de Amparo à Pesquisa do Estado de Goiás (FAPEG), Financiadora de Estudos e Projetos (FINEP), Conselho Nacional de Desenvolvimento Científico e Tecnologico (CNPq) through the scholarship in Research Productivity 1C under the Nr. 304967/2018-1 (C.J) and scholarships in Research Productivity 2 under the Nr. 308753/2022-4 (A.L.M.), National Institute of Photonics (INCT de Fotônica). C.V.T.M., R.F.S. and D.F.L. thank CAPES for their Master scholarships.

**CONFLICT OF INTEREST**

The authors declare no conflict of interest.

**CODE AVAILABILITY**

Our open-source code is available at Ref. [43].

**DATA AVAILABILITY**

The data that support the findings of this study are available together with the Python code at Ref. [43].